\documentclass[preprint, showpacs, floatfix, nofootinbib]{revtex4}
\usepackage{graphicx}
\begin{document}

\title{Evolution of elliptic and triangular flow as a function of beam energy in a hybrid 
model}

\author{J.~Auvinen}
\affiliation{Frankfurt Institute for Advanced Studies (FIAS),
        Ruth-Moufang-Strasse 1, D-60438 Frankfurt am Main, Germany}
\email{auvinen@fias.uni-frankfurt.de}

\author{H.~Petersen}
\affiliation{Frankfurt Institute for Advanced Studies (FIAS),
         Ruth-Moufang-Strasse 1, D-60438 Frankfurt am Main, Germany}
\affiliation{Institut f\"ur Theoretische Physik, Goethe Universit\"at, 
Max-von-Laue-Strasse 1, D-60438 Frankfurt am Main, Germany}
\email{petersen@fias.uni-frankfurt.de}

\begin{abstract}
Elliptic flow has been one of the key observables for establishing the finding of the 
quark-gluon plasma (QGP) at the highest energies of Relativistic Heavy Ion Collider (RHIC) 
and the Large Hadron Collider (LHC). As a sign of collectively behaving matter, one would 
expect the elliptic flow to decrease at lower beam energies, where the QGP is not produced. 
However, in the recent RHIC beam energy scan, it has been found that the inclusive charged 
hadron elliptic flow changes relatively little in magnitude in the energies between 7.7 and 
39 GeV per nucleon-nucleon collision. We study the collision energy dependence of the 
elliptic and triangular flow utilizing a Boltzmann + hydrodynamics hybrid model. Such 
a hybrid model provides a natural framework for the transition from high collision energies, 
where the hydrodynamical description is essential, to smaller energies, where the hadron 
transport dominates. This approach is thus suitable to investigate the relative importance 
of these two mechanisms for the production of the collective flow at different values of 
beam energy. Extending the examined range down to 5 GeV per nucleon-nucleon collision allows 
also making predictions for the CBM experiment at FAIR.
\end{abstract}


\maketitle

\section{Introduction}

In 2010, the RHIC beam energy scan program was launched to study the features of 
the QCD phase diagram and to search for signs of the possible 
first-order phase transition between the confined and deconfined matter
\cite{Kumar:2011de}. The existence of a critical point marking the boundary of cross-over 
and the aforementioned first-order phase transition in the plane of baryochemical potential 
$\mu_B$ and 
temperature $T$ was predicted by lattice calculations \cite{fodor,ejiri,gavai}; it has, however, 
been put to question by the continuum extrapolated results \cite{Aoki:2006we,Endrodi:2011gv} 
which suggest the phase transition remaining cross-over also at large values of $\mu_B$.

Elliptic flow $v_2$ is one of the key observables that supports the formation of a strongly 
coupled quark-gluon plasma at the highest energies of RHIC and the Large Hadron Collider 
(LHC). Thus the naive expectation for $v_2$ in the beam energy scan would be a decrease 
at lower beam energies where the hydrodynamic phase is short or the QGP is not 
created at all. However, the measured inclusive charged hadron elliptic flow $v_2$ 
demonstrates relatively little dependence on the collision energy $\sqrt{s_{NN}}$ between 
7.7 and 39 GeV \cite{Adamczyk:2012ku}.

One possible method for investigating the importance of the hydrodynamical evolution for the 
flow production is the hybrid approach, where one uses a transport model for the 
non-equilibrium phases at the beginning and in the end of a heavy-ion collision event, and 
hydrodynamics for the intermediate hot and dense stage and the phase transition between 
the quark-gluon plasma and hadronic matter. This approach should be applicable 
for a wide range of heavy ion collision energies and thus optimal for studying the beam 
energy dependence of the flow observables down to $\sqrt{s_{NN}}=5$ GeV, an energy reachable 
also at the future heavy ion collisions at FAIR.

\section{Hybrid model}

In this study, a transport + hydrodynamics hybrid model by Petersen {\em et al.} 
\cite{Petersen:2008dd} is utilized. The initial state in this model is produced by the 
Ultrarelativistic Quantum Molecular Dynamics (UrQMD) string / hadronic cascade 
\cite{bass,bleicher}.
The transition to hydrodynamics is done when the two colliding nuclei have passed through 
each other: $t_{\textrm{start}}=\textrm{max}\{2R(\gamma_{CM}^2-1)^{-1/2},0.5 \textrm{ fm}\}$, 
where $R$ represents the nuclear radius and $\gamma_{CM}=(1-v_{CM}^2)^{-1/2}$ 
is the Lorentz factor. The minimum time of 0.5 fm has been determined by the model 
results at the collision energy $\sqrt{s_{NN}}=200$ GeV \cite{Petersen:2010zt}.
At $t_{\textrm{start}}$, the energy-, momentum- and baryon number densities of the particles, 
represented by Lorentz-contracted 3D Gaussian distributions with the width $\sigma=1.0$ fm, 
are mapped onto the hydro grid. Spectator particles are excluded from this procedure and 
propagated separately in the cascade.

The SHASTA algorithm \cite{rischke1,rischke2} is used
to solve the (3+1)-D ideal hydrodynamics evolution equations. The equation of state (EoS) is 
based on a hadronic chiral parity doublet model with quark degrees of freedom, coupled to 
Polyakov loop to include the deconfinement phase 
transition \cite{Steinheimer:2011ea}. It possesses the important feature
of being applicable also at finite baryon densities. At the end of the hydrodynamical 
evolution, the active EoS is changed to the hadron gas EoS, 
so the active degrees of freedom on both sides of the transition hypersurface match
exactly \cite{Steinheimer:2009nn}.

The particlization, i.e. the transition from hydro to transport, is done 
when the energy density $\epsilon$ reaches the critical value $2\epsilon_0$,
where $\epsilon_0=146$ MeV/fm$^3$ is the nuclear ground state energy density. 
The particle distributions are generated according to the Cooper-Frye formula from 
the iso-energy density hypersurface, which is constructed using the Cornelius hypersurface
finder \cite{Huovinen:2012is}. The rescatterings and final decays of these particles are 
then computed in the UrQMD. The final distribution of particles can then be directly 
compared against the experimental data. It has been tested that the hybrid model has a 
reasonable agreement with the experimental data for particle $m_T$ spectra at midrapidity 
$|y|<0.5$ for energies ranging from $E_{\textrm{lab}}=40$ AGeV to $\sqrt{s_{NN}}=200$ GeV 
\cite{Auvinen:2013qia,Auvinen:2013sba}.

\section{Results}

\subsection{Elliptic flow}

In this study, the flow coefficients $v_n$ are computed from the particle momentum 
distributions using the event plane method \cite{poskanzer,ollitrault}. This, together with 
the new implementation of the Cooper-Frye hypersurface finder and particlization, forms the 
core difference compared to previous studies of elliptic flow in the same hybrid approach 
\cite{petersen2009,petersen2010}. 
The primary interest in the following is to see, if the experimentally observed weak 
sensitivity of the elliptic flow on the collision energy is manifested also in the hybrid 
model results.  

\begin{figure}
\centering
\includegraphics[width=5.1cm]{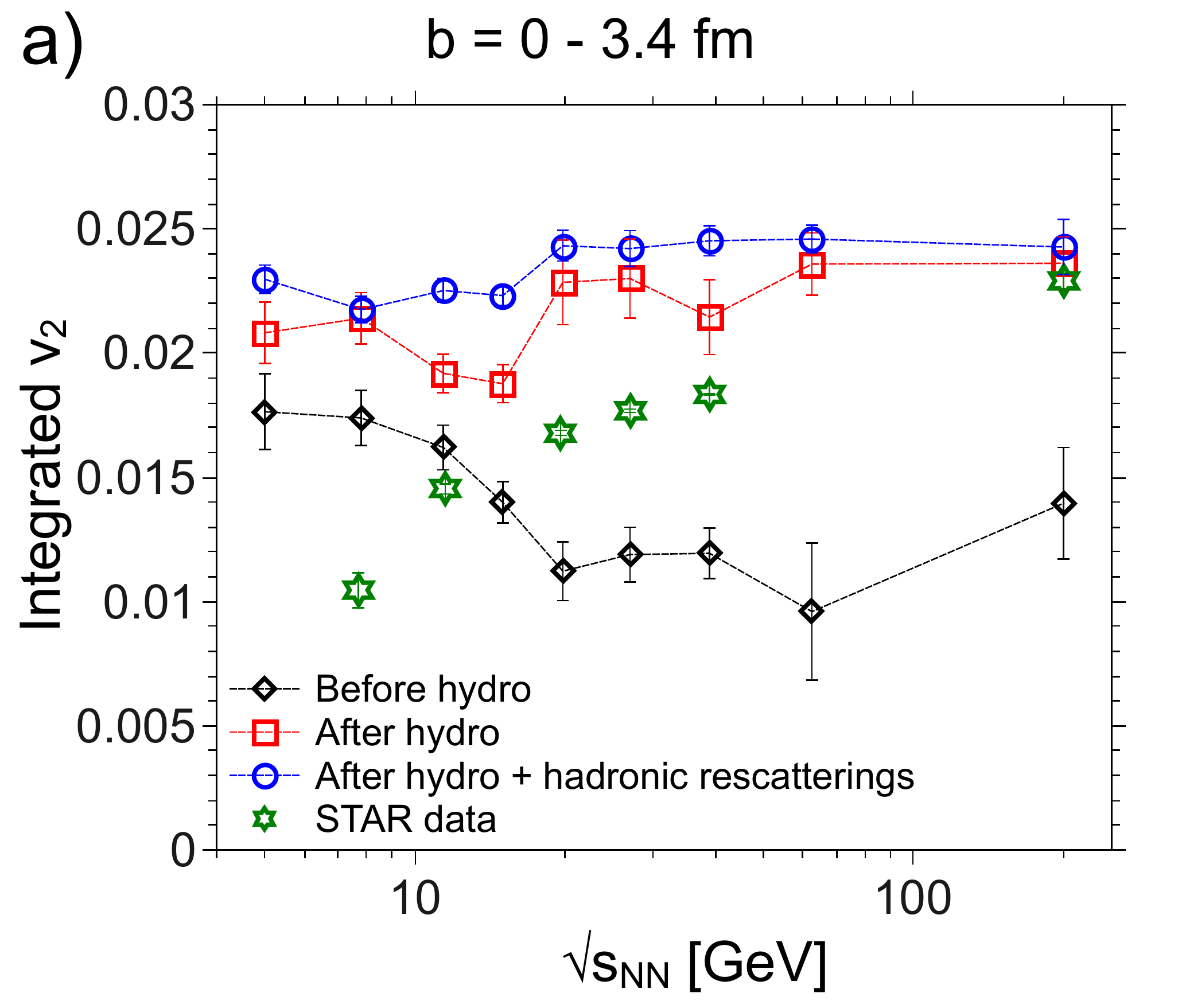}
\includegraphics[width=5cm]{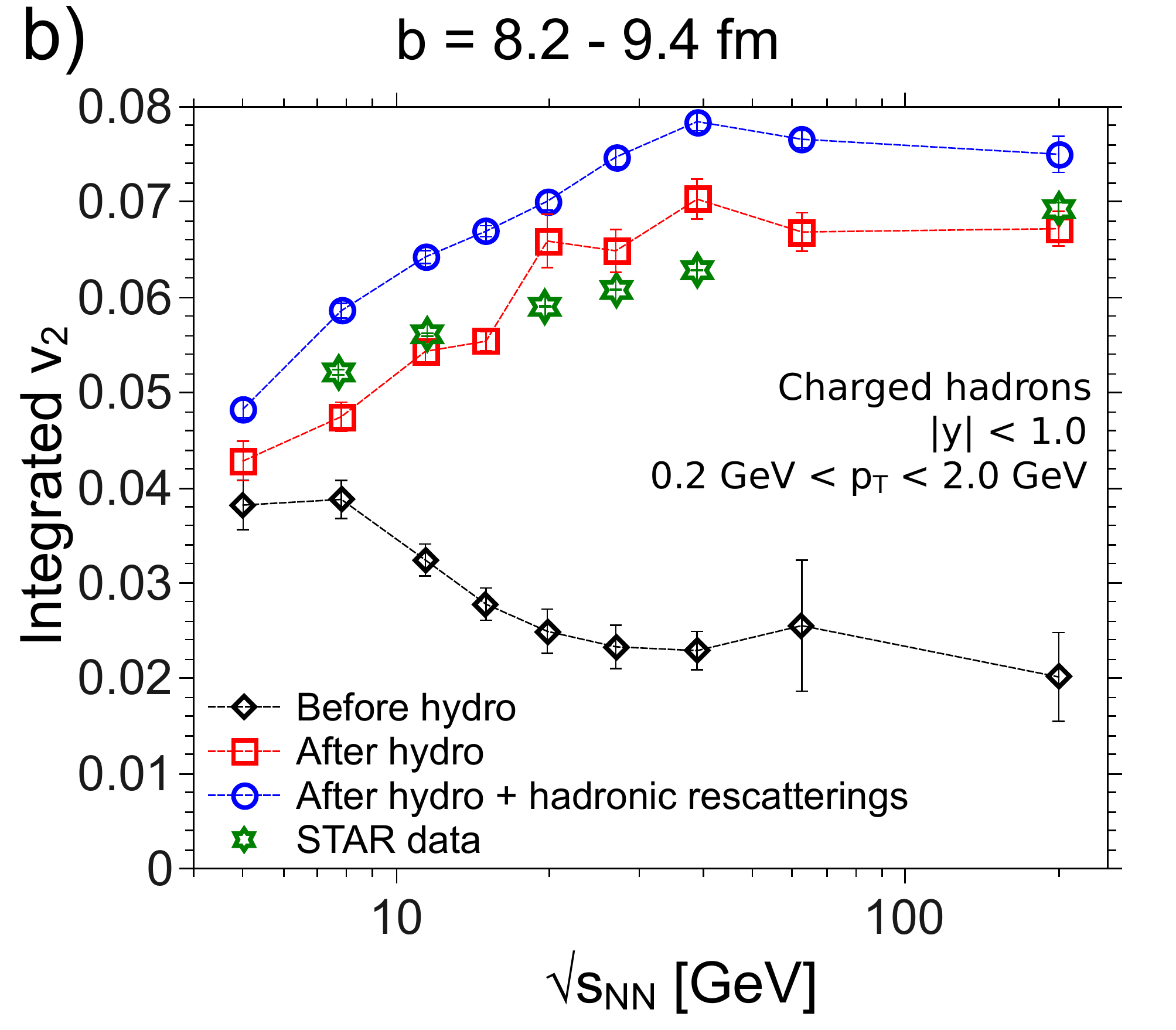}
\includegraphics[width=4.8cm]{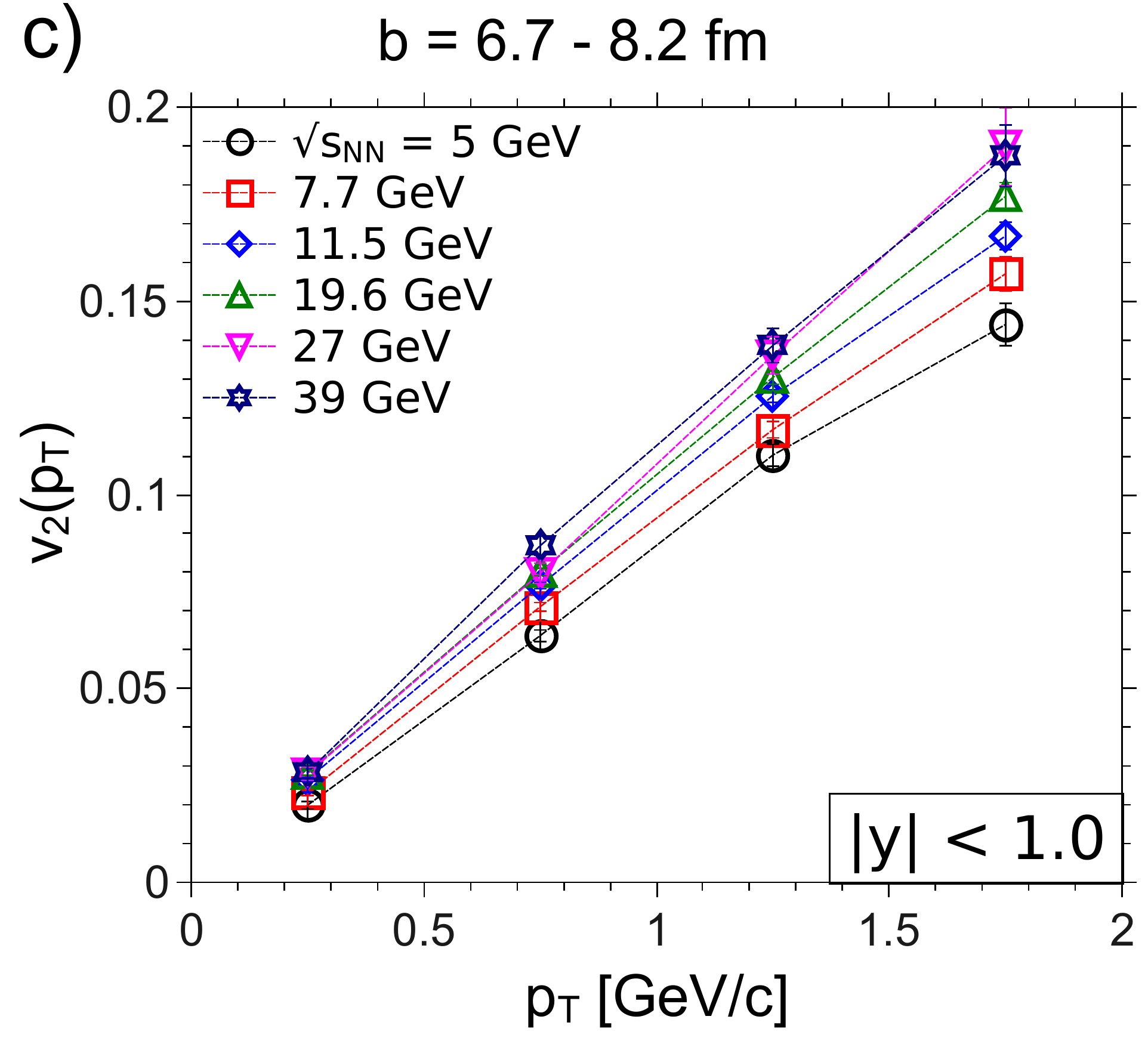}
\vspace{-0.45cm}
\caption{Integrated $v_2$ for $\sqrt{s_{NN}}=5-200$ GeV, at the beginning of hydrodynamical 
evolution (diamonds),
immediately after particlization (squares) and after the full simulation (circles) 
in a) central collisions and b) midcentral 
collisions, compared with the STAR data \cite{Adamczyk:2012ku,Adams:2004bi} (stars). 
c) Differential $v_2$ at midrapidity $|y|<1.0$ for $\sqrt{s_{NN}}=5 - 39$ GeV in impact 
parameter range $b = 6.7-8.2$ fm.}
\label{Figure_v2_phases}
\end{figure}

Figures \ref{Figure_v2_phases}a and \ref{Figure_v2_phases}b show the $p_T$-integrated 
elliptic flow $v_2$ produced in Au+Au -collisions for the $p_T$ range 0.2 - 2 GeV, compared 
with the STAR data for the (0-5)\% and (30-40)\% centrality 
classes. In the model these are respectively represented by the impact parameter intervals 
$b = 0-3.4$ fm and $b = 8.2-9.4$ fm. Figure \ref{Figure_v2_phases}c shows the differential 
$v_2(p_T)$ for $b = 6.7-8.2$ fm, which roughly corresponds to (20-30)\% centrality class.
Figures \ref{Figure_v2_phases}a and \ref{Figure_v2_phases}b also demonstrate the magnitude 
of $v_2$ at three different times: just before the hydrodynamics phase begins, right after 
the hydrodynamics phase has ended and particlization has been done, and after the hadronic 
rescatterings have been performed in the UrQMD (in other words, after the full evolution). 

In the most central 
collisions the effect of the hadronic rescatterings is negligible; in the impact parameter 
range $b = 8.2-9.4$ fm the rescatterings contribute about 10\% on the final result. 
The hydrodynamics also produce very little elliptic flow at $\sqrt{s_{NN}} \leq 7.7$ GeV; 
for the mid-central collisions, $v_2$ is in practice completely produced by the transport 
dynamics, which include resonance 
formation and decay and string excitation and fragmentation processes. These initial 
dynamics, which are often neglected in other hybrid approaches, gain importance 
at lower energies. On the other hand, above $\sqrt{s_{NN}}=19.6$ GeV the hydrodynamic phase 
is clearly the dominant source of $v_2$.

The simulation results overshoot the experimental data for all collision energies. This 
suggests that 
either the viscous corrections should be included, or the energy density value chosen for 
particlization should be higher. In the most central 
collisions below $\sqrt{s_{NN}}=11.5$ GeV, the model appears to produce too much 
flow already at transport phase; here having agreement with the data would require 
modifications in how the pre-equilibrium phase is handled. However, 
for the purposes of this study, the most important feature is the good qualitative 
agreement in the midcentral collisions, as here the flow effects are at their largest. 
Also $v_2(p_T)$ (Fig.~\ref{Figure_v2_phases}c) has relatively weak 
dependence on $\sqrt{s_{NN}}$, which is in accordance with the STAR results.

\subsection{Triangular flow}

Based on the above results, it appears that the hydrodynamically produced elliptic flow 
indeed vanishes, as was the naive expectation, but this is partially compensated by the 
increased flow production in the transport phase and so the observed $v_2$ has only weak 
collision energy dependence. To study this phenomenon further, we do 
the same analysis for another flow observable: the triangular flow $v_3$, which originates 
purely from the event-by-event variations in the initial spatial configuration of the 
colliding nucleons, and is thus largely independent of the collision geometry.

\begin{figure}
\centering
\includegraphics[width=5cm]{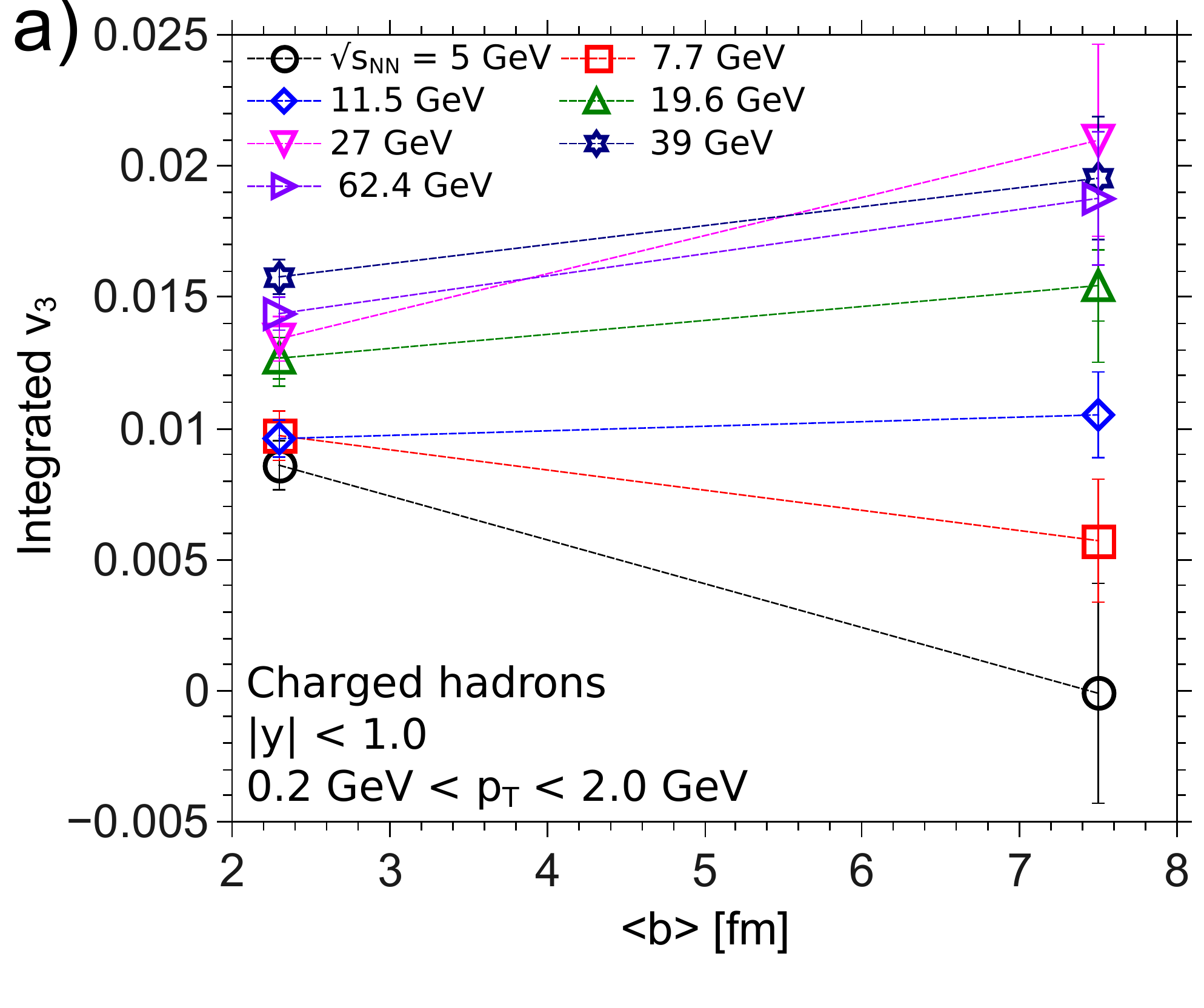}
\includegraphics[width=5cm]{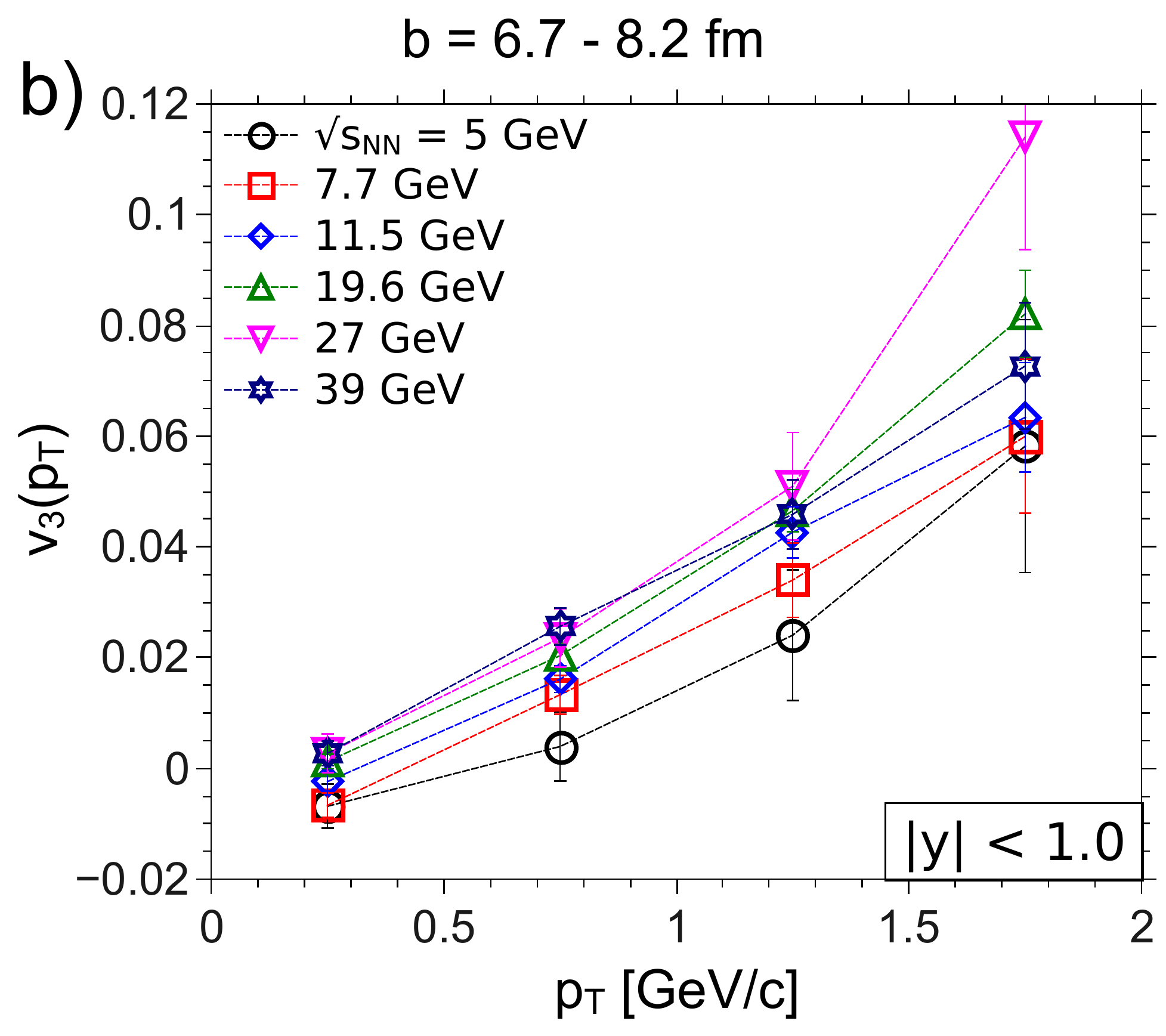}	
\vspace{-0.45cm}
\caption{a) Integrated $v_3$ at midrapidity $|y|<1.0$ in central ($b = 0-3.4$ fm) and 
midcentral ($b = 6.7-8.2$ fm) collisions for collision energies $\sqrt{s_{NN}}=5-62.4$ GeV. 
b) $v_3(p_T)$ in midcentral collisions for $\sqrt{s_{NN}}=5-39$ GeV.}
\label{Figure_v3}
\end{figure}

As illustrated by Figure~\ref{Figure_v3}a, the $p_T$-integrated $v_3$ increases from 
$\approx 0.01$ 
to above 0.015 with increasing collision energy in the most central collisions, whereas 
in midcentrality $b = 6.7-8.2$ fm there is a rapid rise from $\approx 0$ 
at $\sqrt{s_{NN}} = 5$ GeV to the value of $\approx 0.02$ for $\sqrt{s_{NN}} \geq 27$ GeV. 
The collision 
energy dependence is seen also for midcentral $v_3(p_T)$ 
in Fig.~\ref{Figure_v3}b. The energy dependence of $v_3$ in midcentral collisions 
qualitatively resembles the 
hydrodynamically produced $v_2$ in Figure~\ref{Figure_v2_phases}b. Thus 
for the higher flow coefficients, which are more sensitive to viscosity, the transport part 
of the model is unable to compensate for the diminished hydro phase.

\subsection{Effect of initial geometry}

\begin{figure}
\centering
\includegraphics[width=5cm]{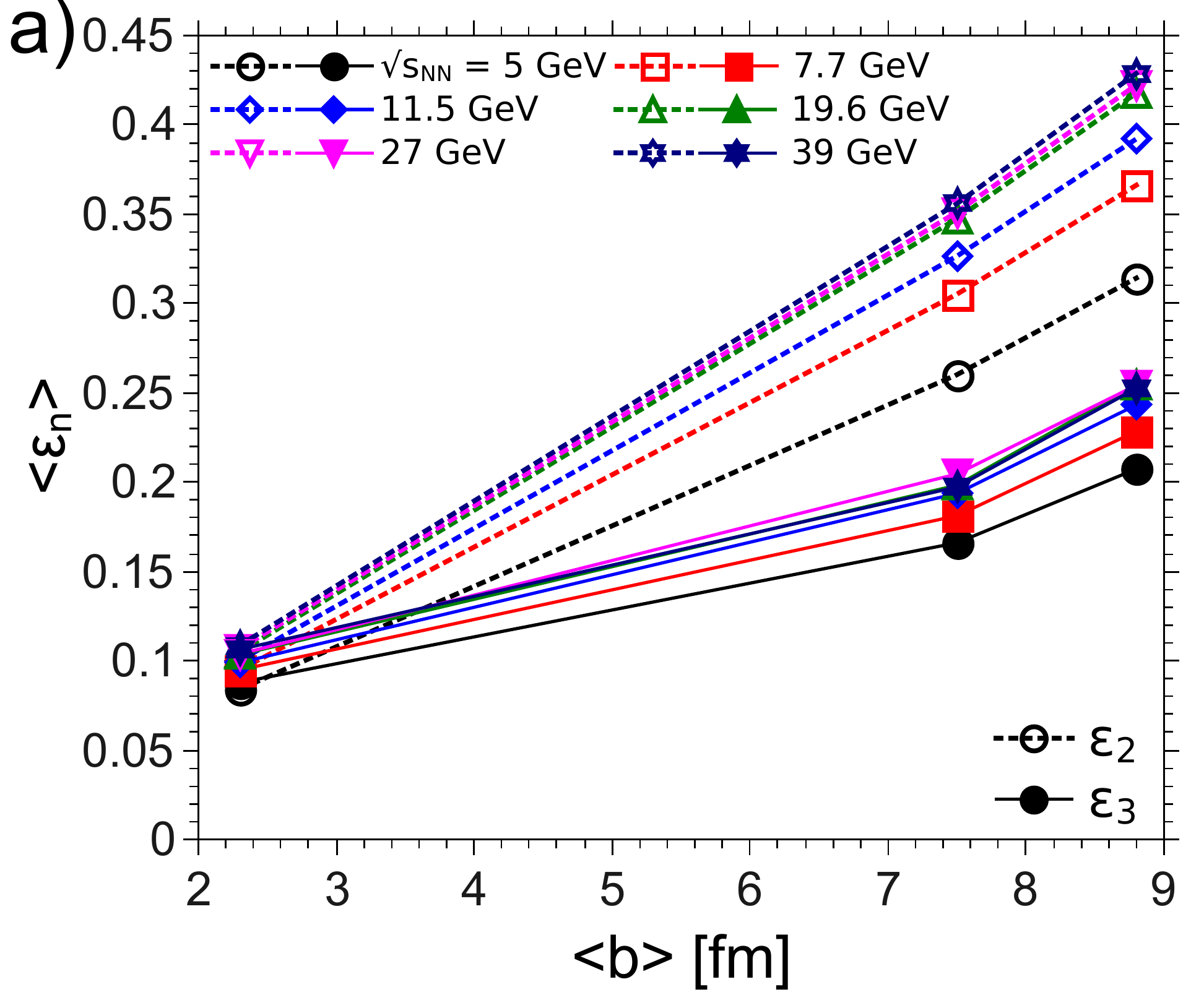}
\includegraphics[width=5.2cm]{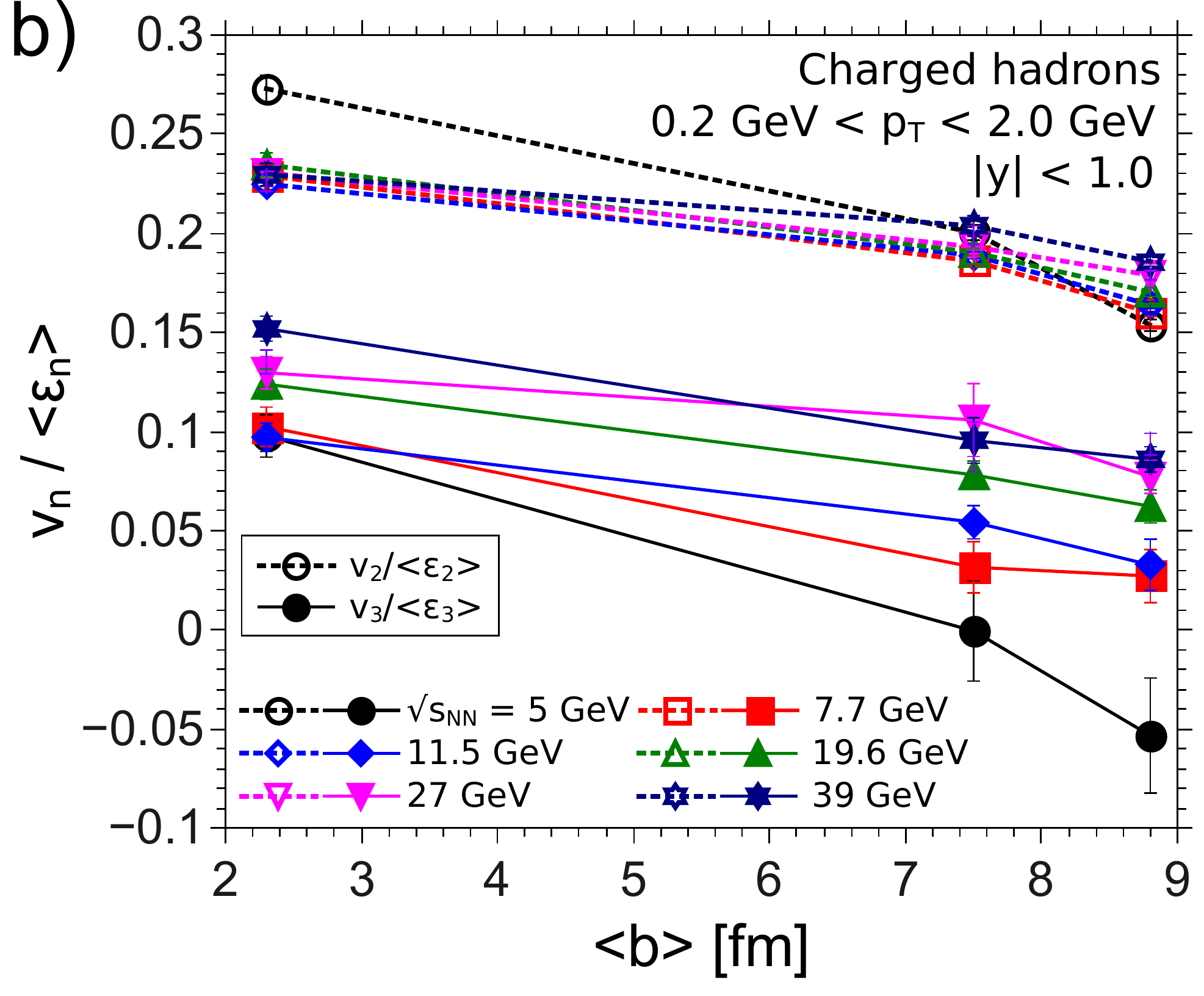}	
\vspace{-0.5cm}
\caption{a) Average eccentricity $\langle \epsilon_2 \rangle$ (open symbols) and 
triangularity $\langle \epsilon_3 \rangle$ (filled symbols)
as a function of average impact parameter $\langle b \rangle$ for the collision 
energy range $\sqrt{s_{NN}}=5-39$ GeV.
b) Scaled flow coefficients $v_2 / \langle \epsilon_2 \rangle$ and 
$v_3 / \langle \epsilon_3 \rangle$ as a function of average impact parameter 
$\langle b \rangle$ for $\sqrt{s_{NN}}=5-39$ GeV.}
\label{Figure_eccentricity}
\end{figure}

Let us then investigate in more detail the effect of initial collision geometry on the flow 
coefficients. Figure~\ref{Figure_eccentricity}a illustrates the collision energy and centrality 
dependencies of the average initial state spatial eccentricity $\langle \epsilon_2 \rangle$ 
and triangularity $\langle \epsilon_3 \rangle$, where the eccentricity and triangularity are 
defined as in \cite{Petersen:2010cw} and calculated at the beginning of 
hydrodynamical evolution $t_{\textrm{start}}$.

The average eccentricity and triangularity are of the same magnitude in the most central 
collisions, where the nuclear overlap region is nearly isotropic. The situation changes 
in mid-central collisions, where, due to the collision 
geometry, $\langle \epsilon_2 \rangle$ is clearly larger than $\langle \epsilon_3 \rangle$. 
The observed dependence on collision energy is largely explained by $t_{\textrm{start}}$, 
which changes rapidly at low energies, from 5.19 fm at $\sqrt{s_{NN}}=5$ GeV to 1.23 fm 
at $\sqrt{s_{NN}}=19.6$ GeV. At low energies there is thus enough time for the 
pre-equilibrium transport to decrease the initial spatial anisotropies.

Figure~\ref{Figure_eccentricity}b shows the coefficients $v_2$ and $v_3$ scaled
with $\langle \epsilon_2 \rangle$ and $\langle \epsilon_3 \rangle$, respectively. 
The relation of the elliptic flow to the initial eccentricity remains largely unchanged 
for the whole collision energy range, while the $v_3$ response to the triangularity 
of the initial state saturates only after 19.6 GeV. This suggests that the hadronic 
medium is too viscous to convert initial state fluctuations into triangular flow, 
and a sufficiently long-living intermediate phase with a low-viscosity fluid is needed for 
the $v_3$ production.

\section{Summary}

We have demonstrated that the experimentally observed behavior of $v_2$ as a function of 
collision energy $\sqrt{s_{NN}}$ can be qualitatively reproduced utilizing a hybrid 
transport + hydrodynamics approach. The diminished hydrodynamical evolution for $v_2$ 
production at lower collision energies is compensated by the pre-equilibrium transport 
dynamics. This compensation does not apply to triangular flow $v_3$, which decreases 
considerably faster, reaching zero in midcentral collisions at $\sqrt{s_{NN}}=5$ GeV. This 
makes $v_3$ the better signal for the formation of quark-gluon plasma in heavy ion 
collisions. However, according to the preliminary STAR data, $v_3$ remains constant 
in central collisions at $\sqrt{s_{NN}}=7.7-27$ GeV \cite{Pandit:QM2012}, 
which suggests that the low-viscous state of nuclear matter is manifested 
at the lower collision energies in greater extent than expected. 
The flow coefficients thus remain interesting observables also for the heavy ion 
collisions at FAIR energies.

\section{Acknowledgements}

The authors acknowledge funding of the Helmholtz Young Investigator Group VH-NG-822. 
This work was supported by the Helmholtz International Center for the Facility for 
Antiproton and Ion Research (HIC for FAIR) within the framework of the 
Landes-Offensive zur Entwicklung Wissenschaftlich-\"okonomischer Exzellenz
(LOEWE) program launched by the State of Hesse. Computational resources have been provided 
by the Center for Scientific Computing (CSC) at the Goethe-University of Frankfurt.

\end{document}